\documentclass[conference]{IEEEtran}
\ifCLASSINFOpdf
   \usepackage[pdftex]{graphicx}
  \graphicspath{{../pdf/}{../jpeg/}}
\else
\fi
\usepackage{amsmath}
\usepackage{listings}

\usepackage[T1]{fontenc}
\usepackage{multirow}
\usepackage{suffix}
\usepackage{url}
 \usepackage{acronym}
\usepackage{booktabs}
\usepackage{empheq}
\usepackage{xcolor}
\usepackage{enumitem}
\usepackage{threeparttable}
\usepackage{array}
\usepackage{hhline}
\usepackage{balance}

\usepackage{cite}
\usepackage[numbers]{natbib}     % this is a better citation system

\usepackage{expl3}
\ExplSyntaxOn
\newcommand\latinabbrev[1]{
  \peek_meaning:NTF . {% Same as \@ifnextchar
    #1\@}%
  { \peek_catcode:NTF a {% Check whether next char has same catcode as \'a, i.e., is a letter
      #1., \@ }%
    {#1., \@}}}
\ExplSyntaxOff

\definecolor{lightpurple}{rgb}{0.8,0.8,1}
\definecolor{codebg}{RGB}{255,255,255}
\definecolor{commentcolor}{RGB}{11,140,11}
\newsavebox{\supbox}% Superscript box
\newcommand{\bsup}{\begin{lrbox}{\supbox}$\tt\scriptstyle}% Superscript begin
\newcommand{\esup}{$\end{lrbox}{}^{\usebox{\supbox}}}% Superscript end
\def\eg{\latinabbrev{e.g}}
\def\ie{\latinabbrev{i.e}}

\newcolumntype{L}[1]{>{\raggedright\let\newline\\\arraybackslash\hspace{0pt}}m{#1}}
\newcolumntype{C}[1]{>{\centering\let\newline\\\arraybackslash\hspace{0pt}}m{#1}}
\newcolumntype{R}[1]{>{\raggedleft\let\newline\\\arraybackslash\hspace{0pt}}m{#1}}

\begin{document}
\title{Towards a Context-Aware  IDE-Based  Meta Search Engine for Recommendation about Programming Errors and Exceptions \vspace{-0.2cm}}


\author{\IEEEauthorblockN{Mohammad Masudur Rahman, Shamima Yeasmin,  Chanchal K. Roy }
\IEEEauthorblockA{Computer Science, University of Saskatchewan, Canada\\
\{mor543, shy942, ckr353\}@mail.usask.ca}
}

%\and
%\IEEEauthorblockN{Iman Keivanloo}
%\IEEEauthorblockA{Computer Science\\
%Concordia University\\
%Montreal, Canada\\
%Email:  i_keiv@encs.concordia.ca}
%}


%Telephone: (800) 555--1212, Fax: (888) 555--1212}


\maketitle

\begin{abstract}
Study shows that software developers spend about 19\% of their time looking for information in the web during software development and maintenance. Traditional web search forces them to leave the working environment (\eg\ IDE) and look for information in the web browser. It also does not consider the context of the problems that the developers search solutions for. The frequent switching between web browser and the IDE is both time-consuming and distracting, and the keyword-based traditional web search often does not help much in problem solving. In this paper, we propose an Eclipse IDE-based web search solution that exploits the APIs provided by three popular web search engines-- Google, Yahoo, Bing and a popular programming Q \& A site, StackOverflow, and captures the content-relevance, context-relevance, popularity and search engine confidence of each candidate result against the encountered programming problems. Experiments with 75 programming errors and exceptions using the proposed approach show that inclusion of different types of context information associated with a given exception can enhance the recommendation accuracy of a given exception. Experiments both with two existing approaches and existing web search engines confirm that our approach can perform better than them in terms of recall, mean precision and other performance measures with little computational cost. 

%The user study conducted with six prospective participants using six exception test cases also reveals a recommendation agreement of 63.33\%. Given all the findings from the conducted experiments, and relevance checking %is subjected to various subjective factors, our proposed approach performs quite well, and it also demonstrates the potential of our proposed idea of IDE-based web search.

\end{abstract}

\begin{IEEEkeywords}
IDE-based search; API mashup; Context-based search; Context code; Cosine similarity;

\end{IEEEkeywords}

% \fi

\IEEEpeerreviewmaketitle

\section{Introduction}
Studies show that up to 80\% of total effort is spent on software maintenance \cite{seahawk}. During development and maintenance of a software product, developers face different programming challenges, and one of the major challenges is software bug fixation. Software bugs are often associated with programming errors or exceptions. Existing IDEs (\eg\ Eclipse, Visual Studio) facilitate to diagnose the encountered errors and exceptions, and developers get valuable information for fixation from the stack traces produced by them. However, the information from the stack trace alone may not help enough in fixation, especially when the developers lack necessary skills or the encountered problems are relatively unfamiliar to them. Thus, for fixation, developers often dig into the world wide web and look for more helpful and up-to-date information. In a study by \citet{twostudy}, developers, in average, spent about 19\% of their programming time in surfing the web for information. \citet{codetrail} conducted a study where they analyzed the events produced by the web browser and the IDE in temporal proximity, and concluded that 23\% web pages visited were related to software development.

Determining the working solution to a programming problem from traditional web search involves trial and error approach in keyword selection, and developers often spend a lot of time to look for such solutions. The search forces them to leave the working environment (\ie\ IDE) and look for the solutions in the web browsers. It also does not consider the context of the problems that they search solutions for. The frequent switching between IDE and the web browser is both distracting and time-consuming, and the keyword-based web search often does not help them much in problem solving. Moreover, checking relevance from hundreds of search results is a cognitive burden on the developers.  

%They also suggest a nice solution to the issue of switching between IDE and browser through visualization of the solution within the IDE
%It facilitates the association of a project source file in the IDE with the StackOverflow posts containing related code examples or exception information

Existing studies focus on integrating commercial-off-the-shelf (COTS) tools into Eclipse IDE \cite{ges}, recommending StackOverflow posts and displaying them within the IDE environment \cite{context, seahawk}, embedding traditional web browser inside the IDE \cite{embed} and so on. \citet{context} propose an IDE-based recommendation system for runtime-exception handling. They extract the question and answer posts from StackOverflow data dump and suggest posts relevant to the encountered exceptions by capturing the exception context from the stack traces generated by the IDE. \citet{seahawk} propose \emph{Seahawk}, an Eclipse IDE plugin, that captures the context (\eg\ source code under editing) of search in terms of several keywords, and recommends StackOverflow posts within the IDE. It also visualizes different components of a recommended post through an embedded and customized web browser for user-friendly use. However, both of the proposed approaches suffer from several limitations.  First, they consider only one source (\eg\ StackOverflow Q \& A site) rather than the whole web for information and thus, their search scope is limited. Second, the developed corpus cannot be easily updated and is subjected to the availability of the data dump. For example, they use the StackOverflow data dump of September 2011, that means, it does not contain the posts generated after September 2011 and therefore, it cannot provide much help or suggestions related to the recently introduced software bugs or errors. Third, they only consider either stack trace or source code under editing as the context of a programming problem for search whereas both of them contain necessary information for fixation. For example, the approach by \citet{context} does not consider the target code that generates the exceptions and therefore, recommends solutions irrespective of the source code context given that the same exception can be generated from different context. Similarly, \emph{Seahawk} \cite{seahawk} cannot answer the programming error or exception related questions as it does not consider the associated stack traces produced by the IDE.

%In this paper, we propose an IDE-based web search solution, \emph{SurfClipse}, to the encountered errors and exceptions which addresses the concerns identified in case of existing approaches. We package the solution as a pluggable Eclipse plugin which collects search results from a remotely hosted web service and displays them within the IDE

In this paper, we propose an IDE-based web search solution, \emph{SurfClipse}, to the encountered errors and exceptions which addresses the concerns identified in case of existing approaches. We package the solution as a pluggable Eclipse plugin prototype which collects search results from a remotely hosted web service \cite{wssurf} and displays them within the IDE. The proposed approach (1) exploits the search and ranking algorithms of three reliable web search engines (\eg\ Google, Bing and Yahoo) and a programming Q \& A site (\eg\ StackOverflow) through the use of API endpoints, (2) provides both a content (\eg\ error or exception message) relevance score and context (\eg\ stack trace, associated source code) relevance score based  ranking on the extracted results of step one, (3) facilitates the most recent solutions, accesses the complete and extensible solution set of StackOverflow and pulls solutions from a number of forums, discussion boards, blogs, programming Q \& A sites and so on, and (4) demonstrates a potential use of web service technology for problem context-aware web search which can be easily leveraged by any IDE of any framework. 

%Fig. \ref{fig:sysdiag} shows our developed plugin for Eclipse IDE which works in two modes-- interactive and proactive. In interactive mode, once developer selects an error or exception (\eg\ Fig. \ref{fig:sysdiag}-(a)), \emph{SurfClipse} captures its context (\eg\ stack trace, associated source code) and generates a web request to the server (\eg\ Fig. \ref{fig:sysdiag}-(d)). The server collects the results from different sources, applies proposed ranking algorithms on the results considering the problem context, and then sends them back to the client plugin. The plugin collects the results and displays them within the IDE (\eg\ Fig. \ref{fig:sysdiag}-(b)). It also facilitates a customized web browser (\eg\ Fig. \ref{fig:sysdiag}-(c)) which provides the real time web browsing experience within the IDE for the developers. In case of proactive mode, the web request is automatically triggered by the plugin when an exception occurs. 

We conduct experiments using \emph{SurfClipse}, and compare with two existing approaches \cite{context, seahawk} and three traditional web search engines with 75 programming errors and exceptions related to Eclipse plugin framework and standard Java applications. The proposed approach recommends correct solutions for 68 (90.66\%) exceptions which essentially outperforms the existing techniques and keyword-based traditional web search approaches in terms of \emph{recall} and other performance measures. We note from the experiments that neither stack trace nor source code alone can determine the complete context of all occurred exceptions, rather their combination represents a more precise context which is likely to produce more relevant results.
 This work is a significantly extended and refined version of our earlier work \cite{masudera}, where we just outlined the idea of an IDE-based meta search engine with limited experiments and validations. 
%In this paper, we propose a distributed model of the IDE-based search, refining the relevance metrics, and adopting parallel processing in the computation for efficient timing. We also conduct a wide range of experiments on the proposed and existing approaches, search engines with a bigger dataset, and the approach is validated with a more extended user study.
%In order to demonstrate the applicability of our proposed approach, we also conduct a user study with six exception-handling test cases involving six prospective participants. We experience 63.33\% agreement between the solutions confirmed or suggested by the participants and the solutions proposed by our approach. 
%Given that relevance checking of a solution to a programming problem is a subjective process and is controlled by different subjective factors, the performance of the proposed approach is quite satisfactory.

The rest of the paper is organized as follows. Section \ref{sec:bg} focuses on the theoretical concepts related our research, Section \ref{sec:model} discusses our proposed system model for IDE-based web search, and Section \ref{sec:metrics} presents our proposed content-based and context-based metrics and algorithms. Section \ref{sec:experiments} discusses about the conducted experiments and experimental results, Section \ref{sec:threats} identifies the possible threats to validity, Section \ref{sec:related} discusses the existing studies related to our research and finally, Section \ref{sec:conclusion} concludes the paper with future works.

\section{Background }
\label{sec:bg}

%\subsection{API Mashups}
%\label{sec:mashup}
%API Mashup can be considered as a technique to combine raw data or services collected from different data sources or web services through the API endpoints, and to produce more useful information or service from them for professional or personal use \cite{mash}. There are three types of API mashups available-- (a) \emph{Business mashups}, where an enterprise combines its own resource or data with external web services to provide an enriched service for developer-customer communication, business requirement management and so on, (b) \emph{Consumer mashups}, where the consumers combines the data and services from different publicly available web services, and (c) \emph{Data mashups}, where data are consumed from similar types of sources and a completely distinct version of data is created from them which is not provided by either of them. In this research, we use \emph{Data mashups} to provide the suggestions to the  reported errors or exceptions by the IDE. We combine the search results against a single query from three web search engines and a programming Q \& A site to provide a more useful and reliable set of results.

%In recent years, more and more web applications have published their APIs and expose their data and services to the developers so that they can reuse those and extend the applications. This technology is playing a vital role in the development of social software and collaborative web applications \cite{mash}. 

\subsection{Cosine Similarity}
\label{sec:cosine}
\emph{Cosine similarity} is a measure that indicates the orientation between two vector spaces with different number of dimensions. It is frequently used in information retrieval to measure the similarity between two text documents where each term is considered as a dimension and each document is considered as a vector of distinct terms. In our research, we often use cosine similarity measure to determine the relevance between the set of query terms and the terms extracted from a candidate result page. We consider each set as a \emph{bag of words} (A collection of words with no fixed order), remove the \emph{stop words} (Insignificant words in a sentence) and then perform \emph{stemming} (A technique to extract the root of a word) on each term which provides their normalized forms. We prepare a combined set of terms, $C$, from two sets and then apply  Equation \eqref{eq:cosine} to calculate the \emph{cosine similarity} score.
\vspace{-0.15cm}
\begin{equation}\label{eq:cosine}
S_{cos}=\frac{\sum_{i=1}^{n}A_i\times B_i}{\sqrt{\sum_{i=1}^{n}A_i^2}\times \sqrt{\sum_{i=1}^{n}B_i^2}}
\end{equation}
%\vspace{-0.1cm}
Here, $S_{cos}$ is the \emph{cosine similarity score}, $A_i$ represents frequency of $i^{th}$ term from $C$ in set A (\eg\ search query), and $B_i$ represents that frequency in set B (candidate result document). This measure values from zero (\ie\ complete lexical dissimilarity) to one (\ie\ complete lexical similarity), and helps to determine the content-relevance between a result page and the search query.

\subsection{Degree of Interest}
\label{sec:doi}
\citet{context} propose \emph{Degree of Interest Score}, which is associated with the reference terms extracted from the stack trace of an occurred exception. They use the score to represent proximity of a reference to the target exception location. In our research, we also leverage this score to determine the relevance between two stack traces. Suppose, a stack trace has \emph{N} method call references, then \emph{Degree of Interest} score for each reference can be calculated using Equation \eqref{eq:doi}.
\vspace{-0.1cm}
\begin{equation}\label{eq:doi}
S_{doi}= 1-\frac{n_{i}-1}{N}
\end{equation}
%\vspace{-0.1cm}
Here, $n_{i}$ represents the position of the reference in stack trace. The score values from zero to one, where zero represents the most distant reference from exception location, and one means the most likely reference that generates the exception.

\section{Proposed Approach for IDE-Based Context-Aware Web Search }

\subsection{Proposed Model for IDE-Based Search}
\label{sec:model}
\begin{figure}[!t]
\centering
\includegraphics[width=2.8in]{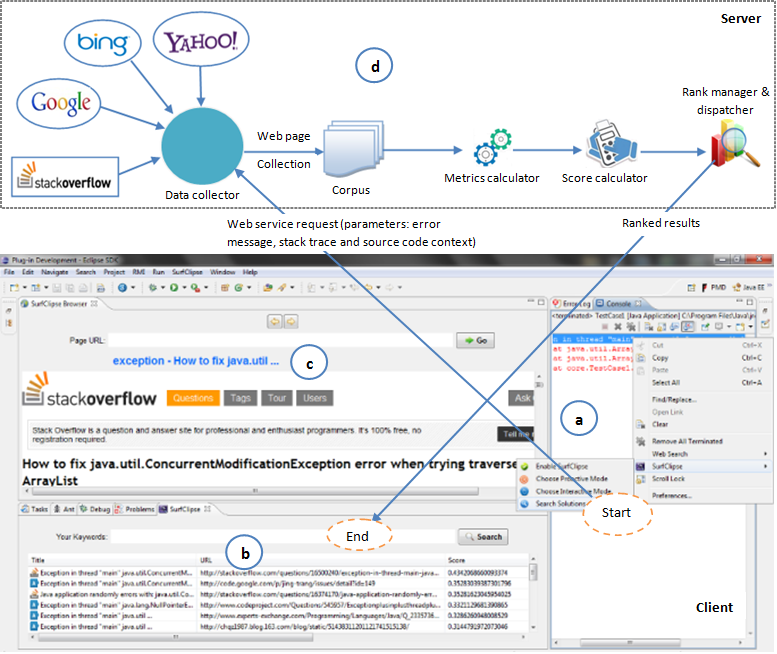}
\caption{Schematic Diagram of the Proposed Approach}
\label{fig:sysdiag}
\vspace{-0.4cm}
\end{figure}

Fig. \ref{fig:sysdiag} shows the schematic diagram of our proposed approach for IDE-based web search. This section discusses the architectural design of our proposed model which includes the working entities, working modes and so on.

\subsubsection{Web Service Provider and Client Plugin}
Our proposed model is based on client-server architecture and it has two major entities-- Eclipse plugin (client) and web service provider (server). They communicate with each other through hyper text transfer protocol and facilitate the search results within the IDE environment. Once the developer selects an encountered exception from \emph{Console View} or \emph{Error Log} in the IDE, the client plugin collects associated context information (\eg\ stack trace and the likely source code causing the exception) and generates a web search request to the service provider \cite{wssurf}. The service provider itself works like a meta search engine, that means, it collects results from multiple search engines against a single query and analyzes them to provide an enriched set of results. In our proposed model, \emph{Data collector} module of the service provider collects results from three state-of-art search engines (Google, Bing and Yahoo) and a programming Q \& A site, StackOverflow, and then accumulates the results to form a \emph{Corpus}. The corpus is generated from about 100-150 results from different sources, and \emph{Metrics calculator} module computes different proposed metrics (Section \ref{sec:metrics}) for each result entry in the corpus. The metrics capture the relevance of the result to the content and context of the query exception. Once metrics are computed, \emph{Score calculator} calculates the final scores of each result entry, and \emph{Rank manager and dispatcher} ranks the results and returns them to the client. The client plugin then captures the results and displays within the IDE in a convenient way. It also facilitates the browsing of the result through a customized web browser widget.

\subsubsection{Plugin Working Modes}
Eclipse plugin in the proposed model works in two modes-- interactive and proactive. In interactive mode (\eg\ Fig. \ref{fig:sysdiag}-(a)), the developer can select the search query by choosing a suitable phrase from the exception stack trace or associated source code, and can make a web search request to the server (\eg\ Fig. \ref{fig:sysdiag}-(d)). Thus, the mode provides a flexible interface (\eg\ Fig. \ref{fig:sysdiag}-(c)) for web search from within the IDE (\eg\ Fig. \ref{fig:sysdiag}-(b)).  In case of proactive mode, the web search request is initiated by the client plugin. In this mode, the plugin assigns a listener to the \emph{Console View} which constantly checks for exception. Once an exception detected, the listener sends error message and other context information to the plugin, and then plugin makes web search request to the service provider. Thus, in this mode, the developer gets rid of the burden of carefully choosing the search query and making the search request manually, and therefore can concentrate on her current tasks without interruption.

\subsubsection{Corpus Development}
Reusing existing data and services to provide an enriched output is an interesting idea, and we use it for the corpus development in our research. We exploit the available API services provided by three popular web search engines (Google, Bing and Yahoo) and a large programming Q \& A site, StackOverflow, to collect the top 30-50 ranked results from each of them against the encountered error or exception, and then use them to develop a corpus dynamically. The idea is to leverage the existing search services and their recommendations to reduce the search scope and produce an effective solution set. Unlike a traditional search engine, which develops an index of all the result pages with some sort of relevance score against a query term, we store necessarily the complete HTML source of each result page as it is parsed and analyzed for relevant stack traces, source code snippets, exception messages and so on in the later phases for metrics calculation.

\subsection{Proposed Metrics}
\label{sec:metrics}

\subsubsection{Search Engine Confidence Score ($S_{sec}$)}
According to \emph{Alexa}\footnote{http://www.alexa.com/topsites, Visited on September, 2013}, one of the widely recognized web traffic data providers, Google ranks first, Yahoo ranks fourth, Bing ranks nineteenth and StackOverflow ranks $67^{th}$ among all websites in the web this year. While these ranks indicate their popularity (\eg\ site traffic) and reliability (\ie\ users' trust) as information service providers, and existing studies show that different search engines perform differently, and even the same search engine performs differently based on the type of search query \cite{gbing, gyahoo}, it is essentially reasonable to think that search results from different search engines of different ranks have different levels of acceptance. To determine the acceptance level of each search service provider, we conduct an experiment with 139 programming errors and exceptions\footnote{http://homepage.usask.ca/$\sim$mor543/query.txt}. We collect the top 10 search results against the  exception query from each search tool and get their \emph{Alexa ranks} \cite{alexa}. Then, we consider the \emph{Alexa ranks} of all result links provided by each search tool and calculate the average rank for a result link provided by them. The average rank for each search tool is then normalized and inversed which provides a value between zero and one, and we consider this value as a heuristic measure of confidence for the search tool. We use Equations \eqref{eq:rnormal} and \eqref{eq:sescore} to get the search engine confidence for a result link.
\vspace{-0.1cm}
\begin{equation}\label{eq:rnormal}
R_{i, normal}=\frac{\bar{R_i}}{\sum_{i=1}^{n}\bar{R_i}}
\end{equation}
\begin{equation}\label{eq:sescore}
S_{i,sec}=\frac{\frac{1}{R_{i,normal}}}{\sum_{i=1}^{n}\frac{1}{R_{i,normal}}}
\end{equation}
%\vspace{-0.1cm}
Here, $\bar{R_i}$ represents the average \emph{Alexa rank} for each search tool results, $R_{i, normal}$ is the normalized version of $\bar{R_i}$ and $S_{i,sec}$ refers to the final confidence score for each search tool based on \emph{Alexa search traffic} statistics. We get a normalized confidence of 0.29 for Google, 0.35 for Bing, 0.36 for Yahoo and 1.00 for StackOverflow given that StackOverflow is a popular programming Q \& A site, which has drawn the attention of a vast programming community (1.9 million\footnote{http://en.wikipedia.org/wiki/Stackoverflow, Visited on September, 2013}) and contains about 12 million questions and answers. The idea is that the occurrence of a result link in multiple search engines against a single query provides the associated confidence scores from the search engines to the link. Thus, if a result occurs in all search provider results, it gets a confidence score of 2.00; however, finally, the scores of all results in the corpus are normalized for practical use.
%if a result link against a single query is found in all three search engines, it gets the search engine score from all three of them which sum to 1.00. StackOverflow has drawn the attention of a vast community of programmers and software professionals, and it also has a relatively better \emph{average Alexa rank}; therefore, the results returned from the programming Q \& A site are provided a search engine score of 1.00.
\subsubsection{Content Matching Score ($S_{cms}$)}
During errors or exceptions, the IDE or Java framework generally issues  notifications from a fixed set of error or exception messages unless the developer handles the exception manually. Therefore, there is a great chance that a search result page titled with an error or exception message similar to the search query would discuss about the encountered problem by the developer, and would contain relevant and useful information for fixation. We propose a metric, \emph{Title to Title Similarity} ($S_{tts}$), that measures the content similarity between the query message and the title of each result page. We use \emph{cosine similarity measure} for this purpose which returns a value between zero (\ie\ completely dissimilar) and one (\ie\ exactly similar). 
As we noted, the result title may not always provide enough information about the content of the exception and therefore, page content needs to be consulted. We consider the stack traces, source code snippets and discussion text extracted from the page content as the legitimate sources of information about the discussed exceptions, and propose two cosine similarity-based metrics-- \emph{Title to Context Similarity} ($S_{tcx}$) and \emph{Title to Description Similarity} ($S_{tds}$). \emph{Title to Context Similarity} score determines the content relevance between query error message and the extracted exception context (\eg\ stack traces, associated code snippets), and \emph{Title to Description Similarity} score denotes the possibility of the occurrence of query error message in the discussion text.

According to \citet{masudir}, terms contained in different parts of the document deserve different levels of attention. For example, a phrase in the page title is more important than a phrase in discussion text to specify the subject matter of the document. We reflect this idea in content matching, and assign different weights to different content similarity scores. We use Equation \eqref{eq:content} to determine the content relevance between the query exception and a result document.
\vspace{-0.10cm}
\begin{equation}\label{eq:content}
S_{cms}=\alpha \times S_{tts}+\beta \times S_{tcx} +\gamma \times S_{tds}
\end{equation}
%\vspace{-0.10cm}
Here, $\alpha, \beta$ and $\gamma$ are the assigned weights to result page title, extracted context (stack trace and code snippet) and discussion text respectively, and they sum to one. Given that the similarity scores are generated from cosine-based measures, \emph{Content Matching Score} always ranges from zero (\ie\ completely irrelevant) to one (\ie\ completely relevant).

\subsubsection{Stack Trace Matching Score ($S_{stm}$)}
To solve the programming errors or exceptions, the associated context such as stack trace generated by the IDE plays an important role. Stack trace contains the error or exception type, system messages and method call references in different source files. In this research, we consider an incentive to the result links containing stack traces similar to that of the query error or exception. We consider both the lexical and structural perspectives of a stack trace, and propose two metrics-- \emph{Lexical Similarity Score} ($S_{lex}$) and \emph{Structural Similarity Score} ($S_{stc}$), to determine the relevance between stack traces. Stack trace generally contains two types of information-- a detailed technical message containing exception name(s) and likely cause(s) of exception, and a list of method call references. We parse the exception names, error message from the first part and extract package, class and method names from each reference in the second part to develop a token set for the stack trace. We use this token set to determine the lexical relevance between two stack traces, and we use \emph{Cosine Similarity Score} for the purpose. It should be mentioned that we do not decompose the camel case tokens into granular levels (\ie\ granularization introduces false positives) in order to perform meaningful similarity checking, which makes the relevance checking effective and useful.

Method call references and their sequence in the stack trace provide important clues to identify the target exception locations and thus, they can help to determine the relevance between two stack traces. We calculate the \emph{Degree of Interest Score} for each reference in the query stack trace using Equation \eqref{eq:doi} and use them to determine \emph{structural relevance} with the candidate stack traces extracted from the result pages. The idea is to determine the occurrence of query method call references in the candidate stack traces. However, complete matching between two references may not be likely and we exploit the idea of \emph{confidence coefficient} proposed by \citet{context}. We get the \emph{Structural Similarity Score} between query stack trace and a candidate stack trace using Equations \eqref{eq:msi} (proposed by \citet{context}) and Equation \eqref{eq:str}.
\vspace{-0.15cm}
\begin{equation}\label{eq:msi}
ms_{i}=S_{doi}\times c_{i}
\end{equation}
\vspace{-0.25cm}
\begin{equation}\label{eq:str}
S_{stc}=\frac{1}{N}\sum_{i=0}^{N}ms_{i}
\end{equation}
%\vspace{-0.15cm}
Here, $ms_{i}$ denotes the matching score between two references, $c_{i}$ refers to confidence coefficient and $N$ represents the number of method call references in the query stack trace. Once both lexical and structural relevance are found, we get the \emph{Stack Trace Matching Score} using the following equation.
\vspace{-0.15cm}
\begin{equation}\label{eq:stm}
S_{st}=\delta \times S_{stc} + \sigma \times S_{lex}
\end{equation}
%\vspace{-0.10cm}
Here, $\delta$ and $\sigma$ denote two heuristic weights assigned to \emph{structural similarity} and \emph{lexical similarity} scores respectively, and they sum to one. \emph{Stack Trace Matching Score} values from zero and one, where zero represents total irrelevance and one represents the complete relevance between two stack traces.
%
%The result pages may contain stack traces which are generated by the IDE due to the same type of error or exception as the selected one. However, they are generated within a different context or different user programs than that of the selected one. Therefore, the complete lexical similarity between the stack traces is not likely and partial similarity with a suitable threshold is a good choice to determine their relevance. \emph{SimHash Algorithm} performs ber for partial similarity matching between two blocks of contents \cite{simhash} and we use it to determine relevance between corresponding stack traces. We extract the stack trace information from the result page through HTML scrapping and apply \emph{SimHash Algorithm} on both stack traces. We get their \emph{SimHash} values and determine the \emph{Hamming distance}. We repeat the process for all result links containing stack traces and use Equation \ref{eq:stscore} to determine their \emph{Stack Trace Scores}.
%\begin{equation}\label{eq:stscore}
%S_{st}=1-\frac{{d_{k}}-\alpha}{max(d_{k})-\alpha}
%\end{equation}
%Here, $d_{k}$ represents the \emph{Hamming distance} between the \emph{Hash values} of each result stack trace and the selected stack trace,$max(d_{k})$ represents the maximum \emph{Hamming distance} found, $\alpha$ represents the minimum \emph{Hamming distance} found and $S_{st}$ refers to the \emph{Stack Trace Score}. 
%The score values from zero to one and indicates the similarity between two stack traces and thus, it can help us to take more informative decision during searching and ranking of results.

\subsubsection{Source Code Context Matching Score ($S_{ccx}$)}
Sometimes, stack trace alone may not provide enough information for the diagnosis of the occurred exception, and the associated code generating the exception needs to be consulted. In programming Q \& A sites, forums and discussion boards, users often post source code snippets related to the exception besides the stack traces for clarification. We are interested to check if the code snippets in the result link are similar to the source code associated with the query exception. This coincidence is possible with the notion that the developers often reuse code snippets from different programming Q \& A sites, forums or discussion boards in their programs directly or with minor modifications. Therefore, a result link containing code snippets similar to the code associated with the query exception is likely to discuss relevant issues that a developer needs to know in order to handle the exception. We consider the code surrounding the exception location in the source file as the source code context of the target exception, and use a \emph{code clone detection technique} \cite{croy} to determine its relevance with the code snippets extracted from the result pages. The idea is to identify the \emph{longest  common subsequence of tokens} between two token sets extracted from two different code snippets. We use Equation \eqref{eq:ccontext} to determine the relevance between the context code associated with query exception and the code snippets extracted from the result page.
\begin{equation}\label{eq:ccontext}
S_{ccx}=\frac{\left | S_{lcs} \right |}{\left |S_{total} \right  |}
\end{equation}
Here, $S_{lcs}$ denotes the longest common subsequence of tokens, and $S_{total}$ denotes the set of tokens extracted from the code blocks considered as the context of the occurred exception. The \emph{Source Code Context Matching Score} values from zero to one.

%We consider three lines before and after the affected line as the source code context of the error or exception and extract the code snippets from result links though HTML scrapping. We apply \emph{SimHash Algorithm} on the code snippets and generate the \emph{SimHash values} and then we use Equation \ref{eq:stscore} to determine \emph{Source Code Context Matching Score} for each result link. The score values from zero to one where one means completely similar and zero means completely dissimilar.

\subsubsection{StackOverflow Vote Score ($S_{so}$)}
StackOverflow, a popular programming Q \& A site with 1.9 million users,  maintains a score for each question, answer and comment posted by the users,  and the score can be considered as a social and technical recognition of their merit \cite{so}. In StackOverflow, a user can up-vote any question or answer post  if she likes something about them, and can also down-vote if the post content seems erroneous, confusing or not helpful. Thus, the difference between up-votes and down-votes from a vast community of technical users, the score of post, is considered as an important metric for evaluation of the quality of the solution posted. In our research, we consider these scores of answer posts against an asked question in the result link from StackOverflow, and calculate \emph{StackOverflow Vote Count} using Equation \eqref{eq:ssum}. Then we normalize the vote count and get \emph{StackOverflow Vote Scoe} using Equation \eqref{eq:soscore}.
\begin{equation}\label{eq:ssum}
SO_{k}=\sum_{\forall p \epsilon P}V_{p}
\end{equation}
\begin{equation}\label{eq:soscore}
S_{so}=\frac{{SO_{k}}- \lambda}{max(SO_{k})-\lambda}
\end{equation}
Here, $SO_{k}$ refers to the StackOverflow vote count for a result page, $V_{p}$ denotes the vote count for a post in the page,  $p$ refers to any post, and $P$ denotes the set of question and answer posts found in a result page. $\lambda$ denotes the minimum vote count, $max(SO_{k}$) represents the maximum vote count and $S_{so}$ is the normalized \emph{StackOverflow Vote Score} for the result link. The score values from zero (\ie\ least significant) to one (\ie\ most significant) and it indicates the relative quality or popularity of the StackOverflow links in the eyes of a large crowd of technical users.

\subsubsection{Search Traffic Rank Score ($S_{str}$)}
The amount of search traffic to a site can be considered as an important indicator of its popularity. In this research, we consider the relative popularity of the result links extracted from different search engines. We use the statistical data from two popular site traffic control organizations-- \emph{Alexa} and \emph{compete} through their provided APIs and get the average popularity rank for each result link. Then, based on these ranks, we provide a normalized \emph{Search Traffic Rank Score} to each result link between zero and one considering minimum and maximum ranks found.

\subsection{Result Scores Calculation}
\label{sec:scores}
The proposed metrics (Section \ref{sec:metrics}) focus on four aspects of each result-- content relevance, context relevance, popularity and search engine confidence, and we consider those aspects for the calculation of final scores. We use \emph{Content Matching Score} of each result page as its \emph{Content Relevance, $(R_{cnt})$} with the encountered exception. In this research, we consider stack trace and the code block triggering the exception as the context of the occurred exception, and use \emph{Stack Trace Matching Score} and \emph{Source Code Context Matching Score} to determine the \emph{Context Relevance, ($R_{cxt}$)} of each result page. Both stack trace and context code carry different levels of significance and we assign two heuristic weights to the matching scores to get the context relevance score.
\begin{equation}\label{eq:contextscore}
R_{cxt}=w_{st}\times S_{stm}+w_{cc}\times S_{ccx}
\end{equation}
Here, $w_{st}$ and $w_{cc}$ are the assigned weights to $S_{stm}$ and $S_{ccx}$ respectively, and they sum to one which gives $R_{cxt}$ normalized, value from zero to one.

\emph{StackOverflow Vote Score} and \emph{Search Traffic Rank Score} are considered as the measures of popularity of each result link from different viewpoints, and they deserve different levels of attention. In the calculation of \emph{Popularity Score}($S_{pop}$) of each result link, we assign two different heuristic weights to these metrics.
\begin{equation}\label{eq:popularity}
S_{pop}=w_{so}\times S_{so}+w_{sr}\times S_{str}
\end{equation}
Here, $w_{so}$ and $w_{sr}$ represent the assigned weights to $S_{so}$ and $S_{str}$ respectively, and they sum to one; this gives $S_{pop}$ a normalized value from zero (\ie\ the least popular) to one (\ie\ the most popular).

Confidence of each result  obtained from the associated search engines can be considered as a support measure for the result link against a search query. We consider \emph{Search Engine Confidence Score} as the confidence of each result. Thus, the four component scores associated with four aspects can be combined using Equation \eqref{eq:final} to get the final score for each result.
\begin{equation}\label{eq:final}
S_{final}=\sum_{\exists w \epsilon W, \exists RS \epsilon (R_{cnt}, R_{cxt}, S_{pop}, S_{sec})}w\times RS   
\end{equation}
Here, $RS$ denotes a measure of content relevance, context relevance, popularity or confidence of a result in the corpus, and $w$ denotes the individual weight (\ie\ importance) associated with each $RS$. We assign a heuristic weight of 0.35 to content-relevance, 0.85 to context-relevance, 0.20 to popularity and 0.10 to the impression of the result link with the search engines.  
We choose these heuristic weights based on our extensive and iterative controlled experiments with a subset of all exceptions, manual analysis on the experimental results, discussion among the authors, and also some helpful ideas from the existing study \cite{context}. While these heuristic values might seem a bit arbitrary, we find the combination to be the best in our experiments to represent the relative importance of different aspects of the final score for a result. Once the final scores are found, the results are sorted in a descending order and the top twenty or thirty of them are returned to the requesting client.

%The proposed search approach considers not only the content and context of the search query but also the popularity of the result as an acceptable solution and the recommendation made by different search tools.

\section{Experiment Design, Results and Validation}
\label{sec:experiments}
\subsection{Dataset preparation}
\label{sec:dataset}
We collect the \emph{workspace logs} of Eclipse IDE from six graduate research students of Software Research Lab, University of Saskatchewan, and extract the exceptions (\eg\ error message, stack trace) occurred during the last six months. We collect a list of 214 stack traces from them. We find that most of the stack traces are duplicate of one another, and we choose 38 distinct stack traces involving 44 \emph{Exception classes}, which are mostly related to Eclipse plugin development framework. To include exceptions related to standard Java application development and prepare a balanced dataset, we choose 37 exceptions from a list of common Java exceptions \cite{scinfo}. Then, we generate some of those exceptions using code examples, and we also perform exhaustive web search to collect the stack traces and source code context associated with those exceptions. We finally get a list of 75 exceptions \cite{scinfo} associated with 75 stack traces \cite{scinfo} and 37 contextual code blocks \cite{scinfo}, which we use as the dataset for different experiments. It should be noted that we cannot collect helpful context codes for the exceptions extracted from the \emph{workspace logs} of the IDE. We collect the most appropriate solutions for the exceptions with the help of different available search engines such as Google, Bing and Yahoo. Given that checking relevance of a solution is controlled by different subjective factors, we select the solution list carefully. First, one of the authors performs exhaustive web searches for two days and collects a potential list of solutions for the exceptions which are shared with other authors. The other authors review the exception information (\eg\ stack trace, code context) and the solution list independently, and provide their feedback about the selection of solution list. Then, the suggestions of all authors are accumulated to finalize a solution set \cite{scinfo} for the exceptions in the dataset.

\subsection{Investigation with Eclipse IDE}
We analyze the features provided by Eclipse IDE in order to find out how often the developers can get necessary help in fixing the encountered exceptions. It is interesting to note that the IDE provides nice debugging tools for them to analyze the exceptions through check-pointing, but they do not help much in platform-level exceptions associated with different runtime libraries and configuration files. For example, if a Java application tries to consume more memory space than alloted, Java framework will issue this exception message-- \emph{java.lang.OutOfMemoryError: Java heap space} and the debugging tools have very little opportunity to suggest about the fixation. There is also an internal web browser widget in recent versions of Eclipse IDE; however, it is  intended for browsing web pages and is defaulted to \emph{Bing Search} which does not consider the context of the encountered exceptions, and thus cannot help the developers much effectively.

\subsection{Search Query Formulation}
Every web search request from Eclipse client plugin has three components-- query for search engines, stack trace and code context. Our proposed solution works in two modes-- interactive and proactive. In case of interactive mode, the developer forms the search query by carefully selecting keywords from the exception message and context information, whereas the plugin is responsible for preparing the search query itself  in case of proactive mode. This section discusses the query formulation techniques for proactive mode.

In this research, we consider both stack trace and the code block likely responsible for the exception as the context for the exception. Therefore, we collect information from the context besides the exception message to generate the search query for the exception. Traditional search engines generally do not allow \cite{longq} or perform poorly for big queries \cite{gbing, gyahoo}, and in order to collect results, we use a sophisticated technique to describe the exception and its context in terms few tokens. We capture the exception message containing exception class name returned by the IDE and collect five class and method names with the highest \emph{Degree of Interests} from the stack trace \cite{context}. We also extract five most frequent method calls and imported class names from the context code using an \emph{ASTParser library}\footnote{http://code.google.com/p/javaparser/} (\ie\ in case of compilable code) and a custom island parser (\ie\ in case of uncompilable code) \cite{seahawk}. Then, we combine the extracted method and class names from both context to generate a unique list of query terms and add the list to the exception message. We note that the exception message itself returned by the IDE is a good descriptor of the exception; however, we filter the message and discard the irrelevant components such as absolute file path, URL and so on. Thus, the filtered exception message and the list describing the context of the exception develop the search query which we use to collect results from different search engines in proactive mode.
\subsection{Performance Metrics}
\label{sec:permet}
Given that our proposed approach is aligned with the research areas of information retrieval and recommendation systems, we use a list of performance metrics from those areas as follows.
%\subsubsection{Mean Average Precision at K (MAPK)}
%While \emph{precision} denotes the fraction of retrieved results that are relevant to the query, \emph{precision at K} calculates \emph{precision} at the occurrence of every relevant document in the ranked list. \emph{Average Precision at K (APK)} averages the precision at K for all relevant results, and we get \emph{Mean Average Precision} for all queries using the following equations.
%\begin{equation}\label{eq:avep}
%APK=\frac{\sum_{k=1}^{D}P_{k}\times rel_{k}}{\left |RR \right |}
%\end{equation}
%\begin{equation}
%MAPK=\frac{\sum_{q\epsilon Q}APK(q)}{\left |Q\right |}
%\end{equation}
%Here, $P_{k}, rel_{k}$ denote the precision at $k_{th}$ result and relevance of $k_{th}$ result respectively, $D$ refers to number of total results, $RR$ is the set of relevant results, $q$ refers to each query, and $Q$ is the set of all queries.
\subsubsection{Mean Precision (MP)}
While \emph{precision} denotes the fraction of retrieved results that are relevant to the query, \emph{Mean Precision} is the average of that measure for all queries.
%\begin{equation}\label{eq:avep}
%APK=\frac{\sum_{k=1}^{D}P_{k}\times rel_{k}}{\left |RR \right |}
%\end{equation}
%\begin{equation}
%MAPK=\frac{\sum_{q\epsilon Q}APK(q)}{\left |Q\right |}
%\end{equation}
%Here, $P_{k}, rel_{k}$ denote the precision at $k_{th}$ result and relevance of $k_{th}$ result respectively, $D$ refers to number of total results, $RR$ is the set of relevant results, $q$ refers to each query, and $Q$ is the set of all queries.

\subsubsection{Mean First False-Positive Position (MFFP)}
First false-positive position (FFP) is the rank of first false-positive result in the ranked list. \emph{Mean First False-Positive Position} measures the average First false-positive position for each query in the query set.
\subsubsection{Mean Reciprocal Rank (MRR)}
\emph{Reciprocal Rank} is the multiplicative inverse of the rank of first relevant result. \emph{Mean Reciprocal Rank} is a statistical measure that averages the \emph{Reciprocal Rank} for each query in the query set.
\subsubsection{Recall (R)} \emph{Recall} denotes the fraction of the relevant results that are retrieved. In our experiments, we consider \emph{recall} as the percentage of the test cases for which the solutions are recommended correctly.

\subsection{Experimental Results on Proposed Approach}
In our experiments, we conduct search with each exception in the dataset and collect the top 30 results. We consider both working modes-- interactive and proactive, and analyze the results using different performance metrics (Section \ref{sec:permet}). Tables \ref{table:expres}, \ref{table:srcomp} and \ref{table:mset} show the results of the experiments conducted on our approach.

Table \ref{table:expres} shows a comparative analysis between \emph{interactive} mode and \emph{proactive} mode of search by the proposed approach. Here, we see that our tool performs relatively better in interactive mode than proactive mode in terms of different performance metrics such as Mean Precision (MP), Mean First False-Positive Position (MFFP), Mean Reciprocal Rank (MRR) and Recall (R). We also note that \emph{proactive version} can recommend correct solutions for 56 exceptions in total whereas \emph{interactive version} can recommend for 68 out of 75 exceptions, which gives a recommendation accuracy of 90.66\% for our approach. Given that formulating search query is one of the decisive factors for the performance of \emph{interactive approach}, we manually select a list of search-friendly keywords from the context of each exception as the search query. The query collects a richer set of initial results from multiple search engines than that of \emph{proactive approach}, where the search keywords are not fine-tuned for search engines.
%Basically, \emph{interactive approach} involves selecting a list of search-friendly keywords chosen from the context of the exception by the developers, and collects a richer set of initial results from multiple search engines than that of \emph{proactive approach}, where the search keywords are not fine-tuned for search engines.

Table \ref{table:srcomp} investigates the impacts of different aspects-- content relevance, context relevance, popularity and search engine confidence of the result link  in the scoring of each result. It shows how the incremental association of different aspects can improve the search results in terms of performance metrics such as Mean Precision (MP), Recall (R) and so on. We note that the proposed approach provides the highest MP and the highest recall when all four aspects are considered during scoring rather than a single aspect such as \emph{content-relevance}. For example,  it performs the best (\eg\ 90.66\%) in terms of \emph{accuracy} (\eg\ TEF (No. of total exceptions fixed), R (\% of exceptions fixed)) when all four dimensions of the result score are considered, which shows the potential of exploiting associated context information besides search query during search.

Table \ref{table:mset} compares the experimental results achieved against two different sets of exceptions-- one with exception messages and stack traces (\eg\ Set A) and the other with exception messages, stack traces and code context (\eg\ Set B). Here, we see that Set B, that considers code context besides stack trace and error message of an exception, achieves higher accuracies (\eg\ 97.29\% and 86.48\%)  than Set A (\eg\ 84.21\% and 63.16\% ) in both working modes. It also gets better results in terms of other performance metrics such as Mean Precision (MP) and Recall (R). Given the findings, all indicate that a combination of context code and stack trace can better specify the context of the exception rather than stack trace only, and thus, by exploiting the context, the proposed approach can recommend solutions more for the set that captures the combination than the one that do not capture.

\begin{table}[!t]
\caption{Results of Experiments on Proposed Approach}
\vspace{-.2cm}
\label{table:expres}
\centering
\resizebox{2.6in}{!}{%
\begin{threeparttable}[b]
\begin{tabular}{l|l|c|c|c}
\hline
\textbf{Mode} & \textbf{Metric} & \textbf{Top 10\tnote{1}} & \textbf{Top 20\tnote{2}} & \textbf{Top 30\tnote{3}}\\
\hline
\multirow{4}{*}{Interactive} &Mean Precision (MP) & 0.1229 & 0.0736 & 0.0538 \\
\hhline{~----}
& MFFP & 1.2400 & 1.2400 & 1.2400 \\
\hhline{~----}
&MRR & 0.4604& 0.4648 & 0.4669 \\
%\hhline{~----}
%&MR& 0.4134 & 0.4765 & 0.5204 \\
\hhline{~----}
&TEF\tnote{4} & 59(75) & 64(75) & \textbf{68}(75) \\
\hhline{~----}
&Recall (R)\tnote{5} & 78.66\% & \textbf{85.33\%} & \textbf{90.66\%} \\
\hline
\hline
\multirow{4}{*}{Proactive} &Mean Precision (MP) & 0.0866 & 0.0529 & 0.0380 \\
\hhline{~----}
&MFFP & 1.2400 & 1.2400 & 1.2400 \\
\hhline{~----}
&MRR & 0.4009 & 0.4048 & 0.4054\\
%\hhline{~----}
%&MR & 0.3510& 0.4070 & 0.4309 \\
\hhline{~----}
&TEF & 51(75) & 55(75) & 56(75) \\
\hhline{~----}
&Recall (R)& 68.00\% & 73.33\% & \textbf{74.66\%}\\
\hline
\end{tabular}
\begin{tablenotes}
\item[1] Metrics for the top 10 results
\item[2] Metrics for the top 20 results
\item[3] Metrics for the top 30 results
\item[4] No. of exceptions fixed
\item[5] Percentage of exceptions fixed
 \end{tablenotes}
\end{threeparttable}
}
\vspace{-.2cm}
\end{table}

\begin{table*}[!t]
\caption{Experimental Results for Different Score Components}
\vspace{-.2cm}
\label{table:srcomp}
\centering
\resizebox{5.6in}{!}{%
\begin{tabular}{l|l|c|c|c||c|c|c}
\hline
\multirow{2}{*}{\textbf{Score Combination}} & \multirow{2}{*}{\textbf{Metric}} & \multicolumn{3}{c||}{\textbf{Interactive}} & \multicolumn{3}{c}{\textbf{Proactive}}\\
\hhline{~~------}
 & & \textbf{Top 10} & \textbf{Top 20} & \textbf{Top 30}& \textbf{Top 10} & \textbf{Top 20} & \textbf{Top 30}\\
\hline
\multirow{3}{*}{Content ($R_{cnt}$)} & MP & 0.0899 & 0.0607 & 0.0481 & 0.0871 & 0.0528 & 0.0371 \\
 %& MR & 0.2797 & 0.3943 & 0.4608 & 0.3096 & 0.3873 & 0.4184\\
 & TEF &43 & 55 & 65 & 46 & 54 & 56\\
 & Recall (R)& 57.33\% & 73.33\% & 86.66\% & 61.33\% & 72.00\% & 74.66\%\\
\hhline{--------}
\multirow{3}{*}{Content ($R_{cnt}$) and Context ($R_{cxt}$)}& MP & 0.11428 & 0.0699 & 0.0514 & 0.0785 & 0.0499 & 0.0376 \\
%& MR & 0.3839 & 0.4403 & 0.4935 & 0.3034 & 0.3628 & 0.4224\\
& TEF &58 & 63 & 66 & 45 & 52 & 55\\
& Recall (R)& 77.33\% & 84.00\% & \textbf{88.00\%} & 60.00\% & 69.33\% & 73.33\%\\
\hhline{--------}
\multirow{2}{*}{Content ($R_{cnt}$), Context ($R_{cxt}$),}& MP & 0.1157 & 0.0699 &0.0519 & 0.0857 & 0.0542 & 0.0381 \\
%& MR & 0.3897 & 0.4463 & 0.4995 & 0.3299 & 0.4090 & 0.4345\\
& TEF &57 & 63 & 66 & 49 & 55 & 56\\
 and Link Popularity ($S_{pop}$) & Recall (R)& 76.00\% & 84.00\% & \textbf{88.00\%} & 65.33\% & 73.33\% & 74.66\%\\
\hhline{--------}
\multirow{2}{*}{Content ($R_{cnt}$), Context ($R_{cxt}$),}& MP & \textbf{0.1229} & 0.0736 & 0.0538 & 0.0886 &0.0529  & 0.0380  \\
%& MR & 0.4134 & 0.4765 & 0.5204 & 0.3510 & 0.4070 & 0.4309\\
& TEF &59 & 64 & \textbf{68} & 51 & 55 & 56\\
 Link Popularity ($S_{pop}$) and Result confidence ($S_{sec}$) & Recall (R)& 78.66\% & \textbf{85.33\%} & \textbf{90.66\%} & 68.00\% & 73.33\% & \textbf{74.66\%}\\
\hline
\end{tabular}
}
\vspace{-.2cm}
\end{table*}

%\multirow{16}{*}{Proactive} & \multirow{4}{*}{$R_{cnt}$}& MAPK & 0.2654 & 0.2660 & 0.2626 \\
%& & MR & 0.3096 & 0.3873 & 0.4184\\
%& & SF &47 & 57 & 59\\
%& & PSF& 56.62\% & 68.67\% & 71.08\%\\
%\hhline{~-----}
%&\multirow{4}{*}{$R_{cnt}, R_{cxt}$}& MAPK & 0.3051 & 0.2990 & 0.2958\\
%& & MR & 0.3034 & 0.3628 & 0.4224\\
%& & SF &46 & 54 & 58\\
%& & PSF& 55.42\% & 65.06\% & 69.88\%\\
%\hhline{~-----}
%&\multirow{4}{*}{$R_{cnt}, R_{cxt}, S_{pop}$}&MAPK & 0.3232 & 0.3159 & 0.3139 \\
%& & MR & 0.3299 & 0.4090 & 0.4345\\
%& & SF &50 & 57 & 59\\
%& & PSF& 60.24\% & 65.06\% & 71.08\%\\
%\hhline{~-----}
%&\multirow{4}{*}{$R_{cnt}, R_{cxt}, S_{pop}, S_{sec}$}&MAPK & 0.3461 &0.3384  & 0.3291 \\
%& & MR & 0.3510 & 0.4070 & 0.4309\\
%& & SF &53 & 58 & 59\\
%& & PSF& 63.85\% & 69.88\% & \textbf{71.08\%}\\

\begin{table}[!t]
\caption{Experimental Results on Multiple Sets}
\vspace{-.2cm}
\label{table:mset}
\centering
\resizebox{2.3in}{!}{%
\begin{threeparttable}[b]
\begin{tabular}{l|l|c|c}
\hline
\textbf{Mode} & \textbf{Metric} & \textbf{Set A\tnote{1} (38)} & \textbf{Set B\tnote{2} (37)}\\
\hline
\multirow{5}{*}{Interactive} &Mean Precision (MP) & 0.0404 & 0.0604 \\
\hhline{~---}
& MFFP & 1.0000 & 1.4864  \\
\hhline{~---}
&MRR & 0.2695 & 0.6697 \\
\hhline{~---}
%&MR& 0.4311 & 0.6263 \\
%\hhline{~---}
&TEF & 32 & 36  \\
\hhline{~---}
&Recall (R)&\textbf{ 84.21\%} & \textbf{97.29\%}  \\
\hline
\hline
\multirow{5}{*}{Proactive} &Mean Precision (MP) & 0.0263 & 0.0450 \\
\hhline{~---}
&MFFP & 1.0000 & 1.4864 \\
\hhline{~---}
&MRR & 0.2563 & 0.5585 \\
%\hhline{~---}
%&MR& 0.3711 & 0.5018 \\
\hhline{~---}
&TEF & 24 & 32 \\
\hhline{~---}
&Recall (R)& \textbf{63.16\%} & \textbf{86.48\%} \\
\hline
\end{tabular}
\begin{tablenotes}
   	\item[1] Contains exception message and stack trace.
	\item[2] Contains exception message, stack trace and code context.
 \end{tablenotes}
\end{threeparttable}
}
\vspace{-.2cm}
\end{table}

\subsection{Experiments with Existing Approaches}
We compare the results of our proposed approach against two existing IDE-based recommendation systems-- context-based recommendation system by \citet{context} and \emph{Seahawk} by \citet{seahawk}. Both of them collect data from StackOverflow data dump and recommend StackOverflow posts taking the current context of the search into consideration. They select suitable tokens from either stack trace or context code to  describe the problem context in the search query, and recommend solutions in a proactive fashion. We implement both of the existing methods and use them for experiments.

To implement the approach proposed by \citet{context}, we extract the exception name from each exception test case and collect 100 top voted StackOverflow posts discussing about that exception, and develop a corpus for the test case. We download the page source of each item in the corpus and create an \emph{Apache Lucene Index} for the corpus. Then, we use the index and \emph{Lucene search engine} to retrieve the relevant posts against the search query associated with the test case. From \emph{Lucene}, we also collect the \emph{retrieval score} based on \emph{Vector Space Model} for each retrieved post. We calculate the \emph{structural score} and \emph{lexical score} of each post considering their stack traces and then normalize them. Finally, we add all three scores for each post to get the final score.

In case of \emph{Seahawk} proposed by \citet{seahawk}, we collect the ten most frequent tokens (\eg\ method reference, class name) from the context code as the search query, and retrieve relevant posts from StackOverflow containing suitable code examples or discussions helpful for current state of coding. Given that \emph{Apache Solr} is a search service provider using \emph{Apache Lucene} as the core search engine, we use \emph{Apache Lucene} to collect relevant results against a search query. Basically, we reuse the previously developed indexes of StackOverflow posts and collect the relevant posts as well as their relevance scores.

Table \ref{table:existing} (left part) shows a comparative analysis between the results of two existing approaches-- \citet{context} and \citet{seahawk}, and our proposed approach. The working principles of the existing approaches are similar to that of \emph{proactive} version of our tool, and therefore, we compare them to the \emph{proactive} version. Here, we can see that both of the existing approaches perform poorly in terms of all performance metrics  compared to our approach. In the best case, they can recommend solutions for 24.00\% and 18.92\% of the exceptions respectively. Given the findings, it is evident that depending on a single information source for exception is not a good choice, and the combination of stack trace and code context is a preferable choice to either any of them for reflecting the  exception context during search.

\begin{table}[!t]
\caption{Common and Unique Results from Search Engines}
\vspace{-.2cm}
\label{table:common}
\centering
\resizebox{3.4in}{!}{%
\begin{tabular}{l|c|c|c|c}
\hline
\textbf{Search Query}&\textbf{Common}&\textbf{Google Only}&\textbf{Bing Only}&\textbf{Yahoo Only}\\
\hline
Content Only & 32 & 09 & 16 & 18\\
\hline
Content and Context & 47 & 09 & 11 & 10\\
\hline
\end{tabular}
}
\vspace{-0.2cm}
\end{table}
\begin{table*}[!t]
\caption{Results of Experiments on Existing Approaches and Search Engines}
\vspace{-.2cm}
\label{table:existing}
\centering
\resizebox{6.7in}{!}{%
\begin{threeparttable}[b]
\begin{tabular}{l|c|c|c|c|c||l|c|c|c|c|c}
\hline
\multicolumn{6}{c||}{\textbf{Proactive Mode}} & \multicolumn{6}{c}{\textbf{Interactive Mode}}\\
\hline
\textbf{Recommender} & \textbf{\#TE\tnote{1}}&\textbf{Metric}&\textbf{Top 10} & \textbf{Top 20} & \textbf{Top 30} & \textbf{Search Engine} & \textbf{\#TE}&\textbf{Metric}&\textbf{Top 10} & \textbf{Top 20} & \textbf{Top 30}\\
\hline
\multirow{3}{*}{\citet{context}} & \multirow{3}{*}{75} & MP &0.0202 & 0.0128 & 0.0085 & \multirow{3}{*}{Google}& \multirow{3}{*}{75} & MP&\textbf{0.1571} & 0.0864 & 0.0580 \\
& & TEF\tnote{2}&15 & 18 & 18 & & & TEF& 57 & 57 & 57 \\
& & R&20.00\% &  24.00\%& \textbf{24.00\%} & & &R & 76.00\%& 76.00\% &\textbf{ 76.00\%}\\
\hhline{------------}
\multirow{3}{*}{Proposed Approach} & \multirow{3}{*}{75}& MP & 0.0886 & 0.0529 & 0.0380 &  \multirow{3}{*}{Bing} & \multirow{3}{*}{75}&MP &  0.1013 & .0533 & 0.0364 \\
& & TEF& 51& 55& \textbf{56} & & &TEF & 55 &  58 & 58 \\
& & R&68.00\% & 73.33\% & \textbf{74.66\%} & & & R& 73.33\% & 77.33\% & \textbf{77.33\%}\\
\hhline{------------}
\multirow{3}{*}{\citet{seahawk}} & \multirow{3}{*}{37} & MP &0.0243 & 0.0135 & 0.0099 & \multirow{3}{*}{Yahoo} & \multirow{3}{*}{75}&MP & 0.0986 & 0.0539 & 0.0369 \\
& & TEF&7  & 7 & 7 & & & TEF& 54 &  57 & 57\\
& & R& 18.92\% & 18.92\% &  \textbf{18.92\%} & & &R& 72.00\% & 76.00\%& \textbf{76.00\%}\\
\hhline{------------}
\multirow{6}{*}{Proposed Approach} & \multirow{6}{*}{37}& \multirow{2}{*}{MP} & \multirow{2}{*}{0.1000}& \multirow{2}{*}{0.0621} & \multirow{2}{*}{0.0450} & \multirow{3}{*}{StackOverflow} & \multirow{3}{*}{75}&MP& 0.0226 & 0.0140 & 0.0097  \\
&  & \multirow{2}{*}{TEF}&\multirow{2}{*}{30} & \multirow{2}{*}{\textbf{32}} & \multirow{2}{*}{\textbf{32}} & & & TEF& 14 &  17 & 17 \\
&  & \multirow{2}{*}{R}& \multirow{2}{*}{81.08\%} & \multirow{2}{*}{86.48\%}  & \multirow{2}{*}{\textbf{86.48\%}} & & & R& 18.66\%& 22.66\% & \textbf{22.66\%} \\
\hhline{~~~~~~------}
& & & & & &\multirow{3}{*}{Proposed Approach} & \multirow{3}{*}{75}&MP& \textbf{0.1229} & 0.0736 & 0.0538 \\
& & & & & & & & TEF&59 &  64 & \textbf{68}\\
& & & & & & & & R&\textbf{78.66\%} & \textbf{85.33\%} & \textbf{90.66\%} \\
\hline
\end{tabular}
\begin{tablenotes}
   	\item[1] No. of exceptions used for the experiment.
	\item[2] No. of total exceptions fixed.
 \end{tablenotes}
\end{threeparttable}
}
\vspace{-.3cm}
\end{table*}

\subsection{Experiments with Existing Search Engines}
We compare the results of our proposed approach against four available search engines-- Google, Bing, Yahoo and StackOverflow. The interactive mode of our approach allows the developer to provide a search query which resembles with working principles of the search engines. We develop search query for each exception using suitable tokens from the context, and use them to collect results from the search engines as well as from the proposed approach. It should be mentioned that the search query is simply used to develop the corpus in case of the proposed approach, and final ranking of the results are determined with the help of ranking algorithms using the automatically extracted context information from the IDE. We collect the top 30 results from each search provider and look for expected solutions identified previously (Section \ref{sec:dataset}). 

Table \ref{table:existing} (right part) shows the comparative analysis of results from different search engines and our proposed approach. Here, we see that existing search engines can recommend solutions for at most 58 (77.33\%) out of 75 exceptions, where the proposed approach can recommend for 68 (90.66\%) exceptions. We also note that \emph{Google} performs slightly better than our approach in terms of Mean Precision (MP), but recommends correct solutions for only 57 exceptions. Given that \emph{selection of appropriate query terms} is an essential precondition for successful search, we conduct another experiment with those search engines using two scenarios--keywords from only exception message, and keywords both from exception message and exception context. Table \ref{table:common} shows the results of those two scenarios. Here, we see that the keyword-based query that considers the exception context, provides more relevant results than the one that does not consider. Therefore, the performance of the traditional search engines is subjected to the selection of search keywords, and the appropriateness of this selection entirely depends on the developer's skill. In our case, we choose the search keywords carefully which provides the better results (\eg\ precision) for Google, but it cannot be taken for granted given the uncertainty in query selection. On the other hand, it is interesting to note that our approach, for the same set of queries, can recommend correct solutions for more exceptions with a little compromise in the precision, and the developers get rid of the burden of choosing appropriate tokens for the context. They can select an encountered exception for search, and the plugin itself captures the problem context to recommend relevant results, whereas Google depends entirely on the developers for the context-based information.

Given the precise results from Google search engine, and the correlation between Mean Precision (MP) and Recommendation Accuracy (\eg\ R) observed at Table \ref{table:srcomp}, one may argue that only Google results should be considered for corpus development in the proposed approach. In our research, we investigate whether such corpus is likely to contain the correct solutions for more exceptions or not. We develop corpus for each of 75 exceptions collecting the top 100 results from Google search API against the selected exception, and apply the proposed ranking algorithms. From Table \ref{table:existing} (right part) we find that the Google corpus-based approach can recommend correctly for at most 57 exceptions. Moreover, Table \ref{table:common} shows that each search provider contains some unique recommendations which cannot be exploited if we consider only one search engine. Therefore, the idea of accumulating search results from multiple search engines for corpus development is promising, and it ensures the maximum \emph{Recall (R)} for our approach by leveraging the existing search services. 

We also investigate into why the proposed approach provides slightly less precise results compared to Google. Given that our approach involves scraping of semi-structured data from the result web page, it may sometimes fail to extract the exception context information properly if the page does not contain the information in the expected tags (\eg\ \emph{code, pre, blockquote}). From our manual analysis with ten cases having the most precise and the least precise recommendations, we find that recommended pages containing context information (\eg\ stack trace, context code snippets) relevant to the query exception are likely to rank higher than those which do not contain such information. In case of the least precise results, the recommended pages contain that information either in an unstructured way which is difficult to extract or they do not contain it at all. Given that the proposed approach emphasizes on the context of the problem discussed in a result page, it does not perform well for those test cases. Therefore, improvement of the context information extraction techniques from web page can help to enhance the \emph{precision} of the proposed approach, which we consider as a scope for future study.
\vspace{-.2cm}
\section{Threats to Validity}
\label{sec:threats}
During the research, we identify a few threats to validity which we discuss in this section. First, the proposed approach still does not provide the search results in real time. Given that the approach involves into scraping web page content for context information about the discussed problem, it takes 30-35 seconds in average to return the recommendations. We applied Java based multi-threading to speed up the computation; however, the approach can be made returning results in real time by more extensive parallelization on the web server and we have already designed it for multi-core systems.

During the experiments, we note that the existing search engines evolve rapidly, especially Google, within days and weeks, and the recommendations from the search engine vary over time for the same query. Therefore, the statistics from the experiments with the search engines are very likely to change. Given that our approach exploits the \emph{live API services} from them, it would also evolve, and it is also subjected to the strength and weaknesses of the search engines. However, adoption of meta search based approach is likely to aggregate the strength and mitigate the weaknesses of each individual search engine as we showed the effectiveness.

Most of the programming errors and exceptions we selected for the experiments are frequently encountered by the developers , and their solutions are also widely discussed in the web. One may argue if the wide availability of those solutions contributes to the better performance of the proposed approach or not. Our approach does not differentiate between frequent and rare programming exceptions, and it returns the relevant recommendations as long as sufficient data are collected from the search engines. However, the approach is subjected to the availability of the appropriate context information (\eg\ stack traces, context code) in the web page for relevance checking.

\section{Related Works}
\label{sec:related}
Existing studies related to our research focus on integrating commercial-off-the-shelf (COTS) tools into Eclipse IDE \cite{ges}, recommending StackOverflow posts and displaying them within IDE environment \cite{context, seahawk}, recommending previously visited web pages \cite{reverb} and open source codes \cite{sourcerer}, embedding traditional web browser inside the IDE \cite{embed} and so on. \citet{ges} integrate Google Desktop Search API into Eclipse environment to facilitate customized search for information within the IDE, which can leverage different software maintenance activities. \citet{context} propose an IDE based recommendation system for runtime exceptions. They extract the question and answer posts from StackOverflow data dump, and suggest posts proactively relevant to the occurred exceptions considering the stack trace information generated by the IDE. In contrast to StackOverflow data dump, our research exploits the existing web search and StackOverflow API services to collect filtered and relevant data from multiple sources, and consider context code, popularity and search engine confidence of the result link besides the exception stack trace. \citet{seahawk} also propose an Eclipse IDE based recommendation system, \emph{Seahawk}, that considers the current state of coding and recommends relevant StackOverflow posts containing code examples and discussions helpful to the coding.  However, it does not consider the stack trace as a component of problem context, and its recommendation may not work for programming exception related issues. \citet{embed} embed a custom code search engine, \emph{Blueprint},  in Eclipse IDE and conduct a user study in the laboratory environment  to investigate whether IDE-based browser can help developer productivity compared to stand-alone web browser. They conclude that the tool helped the developers significantly to write better code and find code examples, and task-specific search interface can greatly influence the web search usage. Our research is related to it in the sense that we also try to address the context-switching issues through IDE based web search features and suitable user interfaces. \citet{reverb} propose \emph{Reverb}, a tool that considers the code under active development within the IDE, and proactively recommends previously visited and relevant web pages from the browsing history. \citet{sourcerer}  propse \emph{Sourcerer}, an open source code search engine that considers both \emph{TF-IDF} and the structural relationships among the code elements to recommend Java classes from 1500 open source projects. Both \citeauthor{reverb} and \citeauthor{sourcerer} exploit lexical and structural features of the source code for recommendation. In our research, we apply similar set of features of the code with the focus on matching local code context of an encountered exception in the IDE against that of the exceptions and programming problems discussed in the web pages for relevant recommendation within the IDE.
\balance

\section{Conclusion and Future Works}
\label{sec:conclusion}
To summarize, we propose a novel IDE-based web search solution that exploits three reliable web search engines and a programming Q \& A site through their API endpoints. The approach facilitates to search for helpful information in the web from within the IDE when the developers face different programming errors and exceptions, and it considers both the problem content and problem context during search. It also considers the popularity and the impression of each result link to different search engines during ranking of the results. We conduct experiments on our approach with 75 programming errors and exceptions related to Eclipse plug-in development framework and standard Java applications. We also conduct experiments on the existing approaches by \citet{context} and \citet{seahawk}, three traditional search engines and StackOverflow with the same dataset. Experiments show that our approach outperforms the existing approaches, search engines and StackOverflow search feature in terms of \emph{recall} and other performance metrics. Experiments also show that inclusion of all types of context information during search can speed up the accuracy of a recommendation system in the IDE. In future, we would attempt to increase the precision of our recommendation system, and validate its applicability with a comprehensive user study.

\bibliographystyle{plainnat}
\footnotesize
\bibliography{mybib}

\begin{thebibliography}{18}
\providecommand{\natexlab}[1]{#1}
\providecommand{\url}[1]{\texttt{#1}}
\expandafter\ifx\csname urlstyle\endcsname\relax
  \providecommand{\doi}[1]{doi: #1}\else
  \providecommand{\doi}{doi: \begingroup \urlstyle{rm}\Url}\fi

\bibitem[ale()]{alexa}
{A}lexa {P}age {R}ank {API}.
\newblock URL \url{http://data.alexa.com/data?cli=10&url=domain.name}.

\bibitem[lon()]{longq}
{W}eb {S}earch {Q}uery.
\newblock URL \url{http://en.wikipedia.org/wiki/Web_search_query}.

\bibitem[sci()]{scinfo}
{S}urf{C}lipse {E}xperiment {D}ata.
\newblock URL \url{http://homepage.usask.ca/~mor543/sc/info}.

\bibitem[wss()]{wssurf}
{S}urf{C}lipse {W}eb {S}ervice.
\newblock URL \url{https://srlabg53-2.usask.ca/wssurfclipse}.

\bibitem[Arif et~al.(2009)Arif, Rahman, and Mukta]{masudir}
A.~Arif, M.M. Rahman, and S.Y. Mukta.
\newblock {I}nformation {R}etrieval by {M}odified {T}erm {W}eighting {M}ethod
  {U}sing {R}andom {W}alk {M}odel with {Q}uery {T}erm {P}osition {R}anking.
\newblock In \emph{Proc. ICSPS}, pages 526--530, 2009.

\bibitem[Bajracharya et~al.(2006)Bajracharya, Ngo, Linstead, Rigo, Dou, Baldi,
  and Lopes]{sourcerer}
S.~Bajracharya, T.~Ngo, E.~Linstead, P.~Rigo, Y.~Dou, P.~Baldi, and C.~Lopes.
\newblock Sourcerer: A {S}earch {E}ngine for {O}pen {S}ource {C}ode
  {S}upporting {S}tructure-{B}ased {S}earch.
\newblock In \emph{Proc. OOPSLA}, pages 25--26, 2006.

\bibitem[Brandt et~al.(2009)Brandt, Guo, Lewenstein, Dontcheva, and
  Klemmer]{twostudy}
Joel Brandt, Philip~J. Guo, Joel Lewenstein, Mira Dontcheva, and Scott~R.
  Klemmer.
\newblock {T}wo {S}tudies of {O}pportunistic {P}rogramming: {I}nterleaving
  {W}eb {F}oraging, {L}earning, and {W}riting {C}ode.
\newblock In \emph{Proc. SIGCHI}, pages 1589--1598, 2009.

\bibitem[Brandt et~al.(2010)Brandt, Dontcheva, Weskamp, and Klemmer]{embed}
Joel Brandt, Mira Dontcheva, Marcos Weskamp, and Scott~R. Klemmer.
\newblock {E}xample-{C}entric {P}rogramming: {I}ntegrating {W}eb {S}earch into
  the {D}evelopment {E}nvironment.
\newblock In \emph{Proc. SIGCHI}, pages 513--522, 2010.

\bibitem[Cordeiro et~al.(2012)Cordeiro, Antunes, and Gomes]{context}
J.~Cordeiro, B.~Antunes, and P.~Gomes.
\newblock {C}ontext-based {R}ecommendation to {S}upport {P}roblem {S}olving in
  {S}oftware {D}evelopment.
\newblock In \emph{Proc. RSSE}, pages 85 --89, June 2012.

\bibitem[Goldman and Miller(2009)]{codetrail}
Max Goldman and Robert~C. Miller.
\newblock {C}odetrail: {C}onnecting {S}ource {C}ode and {W}eb {R}esources.
\newblock \emph{J. Vis. Lang. Comput.}, 20\penalty0 (4):\penalty0 223--235,
  August 2009.

\bibitem[Kumar and Prakash(2009)]{gyahoo}
B.T.~S. Kumar and J.N. Prakash.
\newblock {P}recision and {R}elative {R}ecall of {S}earch {E}ngines: {A}
  {C}omparative {S}tudy of {G}oogle and {Y}ahoo.
\newblock \emph{J. Lib. and Info. Mgmt.}, 38\penalty0 (1):\penalty0 124--137,
  2009.

\bibitem[Nasehi et~al.(2012)Nasehi, Sillito, Maurer, and Burns]{so}
S.M. Nasehi, J.~Sillito, F.~Maurer, and C.~Burns.
\newblock {W}hat {M}akes a {G}ood {C}ode {E}xample?: {A} {S}tudy of
  {P}rogramming {Q} \& {A} in {S}tack{O}verflow.
\newblock In \emph{Proc. ICSM}, pages 25--34, 2012.

\bibitem[Ponzanelli et~al.(2013)Ponzanelli, Bacchelli, and Lanza]{seahawk}
Luca Ponzanelli, Alberto Bacchelli, and Michele Lanza.
\newblock Seahawk: {S}tack {O}verflow in the {IDE}.
\newblock In \emph{Proc. ICSE}, pages 1295--1298, 2013.

\bibitem[Poshyvanyk et~al.(2007)Poshyvanyk, Petrenko, and Marcus]{ges}
D.~Poshyvanyk, M.~Petrenko, and A.~Marcus.
\newblock {I}ntegrating {COTS} {S}earch {E}ngines into {E}clipse: {G}oogle
  {D}esktop {C}ase {S}tudy.
\newblock In \emph{Proc. IWICSS}, pages 6--, 2007.

\bibitem[Rahman et~al.(2013)Rahman, Yeasmin, and Roy]{masudera}
M.M. Rahman, S.~Yeasmin, and C.K. Roy.
\newblock {A}n {IDE}-{B}ased {C}ontext-{A}ware {M}eta {S}earch {E}ngine.
\newblock In \emph{Proc. WCRE}, pages 467--471, 2013.

\bibitem[Roy and Cordy(2008)]{croy}
C.K. Roy and J.R. Cordy.
\newblock {NICAD}: {A}ccurate {D}etection of {N}ear-{M}iss {I}ntentional
  {C}lones {U}sing {F}lexible {P}retty-{P}rinting and {C}ode {N}ormalization.
\newblock In \emph{ICPC}, pages 172--181, 2008.

\bibitem[Sawadsky et~al.(2013)Sawadsky, Murphy, and Jiresal]{reverb}
N.~Sawadsky, G.C. Murphy, and R.~Jiresal.
\newblock Reverb: Recommending code-related web pages.
\newblock In \emph{Proc. ICSE}, pages 812--821, 2013.

\bibitem[Usmani et~al.(2012)Usmani, Pant, and Bhatt]{gbing}
T.~Usmani, D.~Pant, and A.~K. Bhatt.
\newblock A {C}omparative {S}tudy of {G}oogle and {B}ing {S}earch {E}ngines in
  {C}ontext of {P}recision and {R}elative {R}ecall parameter.
\newblock \emph{J. CSE}, 4\penalty0 (1):\penalty0 21--34, 2012.

\end{thebibliography}

\end{document}